\documentclass[]{article}

\newtheorem{theorem}{Theorem}
\newtheorem{lemma}[theorem]{Lemma}
\def\squareforqed{\hbox{\rlap{$\sqcap$}$\sqcup$}}
\def\qed{\ifmmode\squareforqed\else{\unskip\nobreak\hfil\penalty50\hskip1em\null\nobreak
\hfil\squareforqed\parfillskip=0pt\finalhyphendemerits=0\endgraf}\fi\vspace*{4mm}}
\def\proof{\noindent{\it Proof.\ \ \ }}

\title{$k$-Connectivity in the Semi-Streaming Model\thanks{Supported by the DFG Research Center {\sc Matheon} ``Mathematics for key technologies" in Berlin}}
\author{Mariano Zelke\thanks{tel.:+49-30-2093 3196, fax:+49-30-2093 3191}
\\ \small Humboldt-Universit\"at zu Berlin,\\\small Institut f\"ur Informatik,\\\small 10099 Berlin\\ zelke@informatik.hu-berlin.de}
\date{}

\begin{document}

\maketitle

\begin{abstract}
\noindent We present the first semi-streaming algorithms to determine $k$-connectivity of an undirected graph with $k$ being any constant. The semi-streaming model for graph algorithms was introduced by Muthukrishnan in 2003 and turns out to be useful when dealing with massive graphs streamed in from an external storage device.

Our two semi-streaming algorithms each compute a sparse subgraph of an input graph $G$ and can use this subgraph in a postprocessing step to decide $k$-connectivity of $G$. To this end the first algorithm reads the input stream only once and uses time ${\cal O}(k^2n)$ to process each input edge. The second algorithm reads the input $k+1$ times and needs time ${\cal O}(k+\alpha(n))$ per input edge. Using its constructed subgraph the second algorithm can also generate all $l$-separators of the input graph for all $l<k$.
\end{abstract}

\ \\{\bf Keywords:} graph, semi-streaming algorithm, connectivity, vertex connectivity, separator

\section{Introduction}

\textbf{Semi-Streaming Model.} In the recent years the computational model of streaming algorithms has gained popularity, not only because of its interesting theoretical implications but also due to its usefulness in practice. Real-world applications are facing an increasing amount of data, needing the ability to deal with massive amounts of information. Examples vary from oceanographic and atmospheric data to the huge databases of data warehousing. It is common that the input data size of this kind can easily reach terabytes or petabytes. Thus the traditional approaches of algorithms having random access to the input are not useful here. On the contrary it cannot be taken for granted that the whole input is available in the memory of the algorithm, it is rather stored on disk or tape. For developing time-efficient algorithms working on these storage devices it is reasonable to assume the input of the algorithm (which is the output of the storage devices) to be a sequential stream. While tapes produce a stream as their natural output, disks reach much higher output rates when presenting their data sequentially in the order it is stored.

This is where streaming algorithms are placed in position. They provide a computational model useful for dealing with large amounts of data stored in external devices. In the classical \emph{data stream model} \cite{HenzingerRaghavanRajagopalan}, \cite{Muthukrishnan} the input data can only be accessed sequentially as a data stream. The streaming algorithm has to process this input using a working memory that is small compared to the length of the input stream. In particular the algorithm is unable to store the whole input and therefore has to make space-efficient summarizations of the input according to the query to be answered.

Much of the previous work in the area of streaming models is focused on generating statistical values for a stream of input elements. There are streaming algorithms approximating frequency moments \cite{AlonMatiasSzegedy}, computing histograms \cite{GuhaKoudasShim} and wavelet decompositions \cite{GilbertKotidisMuthukrishnan}. For a comprehensive overview the reader is referred to \cite{Muthukrishnan} and the references therein.

Real-world applications often deal with data modeled as a graph $G(V,E)$ composed of vertices $V$ and edges $E$. One example is the call graph of telecommunication providers modeling the users as vertices and the telephone calls as edges between them. A second example is the structure of the WWW where pages are vertices and links correspond to edges. Both are massive graphs and answering queries on these graphs means to solve graph theoretical problems on a huge amount of input.\\ 
\indent The traditional streaming model \cite{Muthukrishnan} restricts an algorithm on a graph with $n$ vertices to a memory size of $o(n)$ bits. That does not suffice even to solve basic graph problems \cite{FeigenbaumKannanMcGregorSuriZhang05}. Therefore Muthukrishnan \cite{Muthukrishnan} proposed the \emph{semi-streaming model} for handling graph issues in the context of streaming: Given a graph $G(V,E)$, $n=|V|$ and $m=|E|$, a \emph{semi-streaming} algorithm is presented an arbitrary order of the edges of $G$ as a stream. The algorithm can only access this input sequentially in the order it is given; it might process the input stream several times. The algorithm has a working memory consisting of ${\cal O}(n\cdot\mbox{polylog}\,n)$ bits, thus there is space to store the vertices but not enough to store the edges of $G$ if $G$ is a dense graph, i.e., $n=o(m)$. 

There have been some successful considerations of graph problems in the semi-streaming model. In \cite{FeigenbaumKannanMcGregorSuriZhang04} a semi-streaming algorithm is given for testing if a graph is connected and for creating a bipartition of the edges or stating that there is not any. In this paper a semi-streaming algorithm is presented that creates a minimum spanning tree of a weighted graph, as well as one that calculates all cut-vertices of a graph. There are approaches in \cite{FeigenbaumKannanMcGregorSuriZhang04}, \cite{McGregor} to approximate a maximum matching in unweighted and weighted graphs. In \cite{FeigenbaumKannanMcGregorSuriZhang05} the authors use the idea of a spanner to develop approximations for all-pair shortest paths, diameter and girth of a graph.\\
\indent On the other hand there are some results showing the limits of the semi-streaming model. We just name two examples. Testing connectivity of a directed graph is not possible in the semi-streaming model \cite{FeigenbaumKannanMcGregorSuriZhang04} and for general graphs a breadth-first search tree cannot be created in a constant number of passes over the input \cite{FeigenbaumKannanMcGregorSuriZhang05}.

\textbf{\boldmath{$k$}-Connectivity.} The notion of \emph{$k$-connectivity} of a graph arises for example by looking at telecommunication networks. These networks have to be robust, even in the case of failures a user of this network should be guaranteed to reach every other user. On the network modeled as a graph, routers and users being the vertices, cables between them being the edges, we could ask how many vertices may fail such that the network is still serving a connection between every pair of users. A graph $G(V,E)$ is said to be \emph{$k$-vertex connected} or \emph{$k$-connected} if after the removal of any $k-1$ vertices $G$ is still connected, that is, it contains a path between each pair of vertices. As a classical topic of graph theory this problem has been extensively studied in both the directed and undirected case, see \cite{Schrijver} for an overview. We only consider the undirected case throughout this paper and we can find the largest $k$ such that a given undirected graph is $k$-connected using a variety of algorithms for example one due to Gabow \cite{Gabow} which runs in time ${\cal O}(n+\min\{k^{5/2}, kn^{3/4}\}kn)$.

\textbf{\boldmath{$k$}-Connectivity in the Semi-Streaming Model.} The situation in the semi-streaming model is quite different. So far only semi-streaming algorithms for specifying $k$-connectivity for $k\le 4$ are known. For $1$-connectivity, which is just connectivity, in \cite{FeigenbaumKannanMcGregorSuriZhang04} a semi-streaming algorithm is given that needs only one pass over the input stream and processes each input edge in time ${\cal O}(\alpha(n))$, where $\alpha(n)$ is the extremely slowly growing inverse of the Ackermann function \cite{Tarjan}. For $k=2,3,4$ the authors of \cite{FeigenbaumKannanMcGregorSuriZhang05} present an adoption of a sparsification technique of \cite{EppsteinGalilItalianoNissenzweig}. That leads to semi-streaming algorithms for testing $2$- and $3$-connectivity in time ${\cal O}(\alpha(n))$ per edge and for testing $4$-connectivity in time ${\cal O}(\log n)$ per edge. These approaches can also be used to identify $l$-separators of $G$ for $l<k$. An $l$-separator of a graph $G$ is a set of $l$ vertices whose removal leaves a graph with more connected components than $G$. However, there is no semi-streaming algorithm determining if a given graph is $k$-connected for any constant $k>4$, not to mention one to find $l$-separators for constant $l>3$.

\textbf{Our Contributions.} In this paper we present the first two semi-streaming algorithms for determining if a given graph is $k$-connected for $k$ being an arbitrary constant. The first algorithm is an adoption of an online algorithm developed in \cite{CheriyanThurimella91} and \cite{CheriyanKaoThurimella93}. It runs over the input only once and takes time ${\cal O}(k^2n)$ to process each input edge. The second algorithm has a faster processing time per input edge of ${\cal O}(k+\alpha(n))$ but needs to read the input stream $k+1$ times. It is based on results in \cite{CheriyanKaoThurimella93}.\\
\indent Both algorithms utilize the idea of a \emph{certificate}. For a graph $G$ a certificate for $k$-connectivity is a subgraph of $G$ such that the certificate is $k$-connected if and only if $G$ is $k$-connected. While reading the input the presented algorithms both construct a certificate that does not exceed the memory limitations of the semi-streaming model. Thus the algorithms can memorize their certificates and can make use of them to determine $k$-connectivity of the input graph in a postprocessing step without any further input.\\
\indent Moreover the certificate of the second algorithm can be used not only to test for $k$-connectivity but even for computing all $l$-separators of a given graph for every $l<k$.


\section{Preliminaries and Definitions}
We denote by $G$ a graph $G(V,E)$ with vertex set $V$ and edge set $E$. Let $n=|V|$ and $m=|E|$ be the number of vertices and the number of edges respectively. Throughout the whole paper $G$ is an undirected, unweighted graph without multiple edges or loops.

We call two vertices \emph{connected} if there is a path between them. A graph $G$ is connected if any pair of vertices in $G$ is connected, a \emph{connected component} of $G$ denotes an induced subgraph $C$ of $G$ such that $C$ is connected and maximal.\\
\indent Given a positive integer $k$, a graph $G$ with at least $k+1$ vertices is said to be \emph{$k$-vertex connected} or \emph{$k$-connected} if the removal of any $k-1$ vertices leaves the graph connected.\\
\indent A subset $S$ of the vertices of $G$ we call an \emph{$l$-separator}, if $l=|S|$ and the graph obtained by removing $S$ and all edges incident to $S$ from $G$ has more connected components than $G$.\\
\indent For two distinct vertices $x,y$ in $G$ we call two paths between $x$ and $y$ \emph{vertex-disjoint} if they are internally vertex-disjoint, that is, have only the endpoint $x$ and $y$ in common. Using that we name $\kappa(x,y)$ the \emph{local connectivity} between $x$ and $y$, being the maximum number of vertex-disjoint paths between $x$ and $y$ in $G$. 

For any property $\cal{P}$ and graph $G$ we define a subgraph $G'=(V,E')$, $E'\subseteq E$, to be a \emph{certificate} for $G$ in the case that $G$ has property $\cal{P}$ if and only if $G'$ has property $\cal{P}$.  Thus a certificate for $k$-connectivity of $G$ is a subgraph $G'$ of $G$ such that $G$ is $k$-connected if and only if $G'$ is $k$-connected. A certificate $G'$ is said to be a \emph{sparse certificate} if $G'$ has a linear number of edges, that is, $|E'|={\cal O}(n)$.

A \emph{graph stream} of a graph $G$ is a sequence of the $m$ edges of $G$ in arbitrary oder. There is no restriction on the order of the edges, for example it is not required that all edges incident with a vertex are grouped together in the sequence. If we consider a graph stream as an input we mean that the edges are revealed one at a time. A \emph{semi-streaming graph algorithm} computes over a graph stream as an input and is allowed to use a space of at most ${\cal O}(n\cdot\mbox{polylog}\,n)$ bits. The algorithm may access the input stream for $P$ passes in a sequential one-way order and use time $T$ to process each single edge.

At some places in the paper we use the function \emph{$\alpha(n)$} which is the inverse of the extremely quickly growing Ackermann function. Therefore $\alpha(n)$ is very slowly growing \cite{Tarjan}.


\section{Testing \boldmath{$k$}-Connectivity in the Semi-Streaming Model}
In this section we present an algorithm to test a graph $G$ for $k$-connectivity in the semi-streaming model for $k$ being an arbitrary positive constant. Our approach uses the concept of sparse certificates. In the next two subsections we develop two different semi-streaming algorithms $A_1$ and $A_2$, each of them computing a sparse certificate for $k$-connectivity of a given graph $G$. 

The basic idea for both algorithms is the same: While  processing the graph stream of $G$ a sparse certificate is built up, $C(A_1)$ by $A_1$ and $C(A_2)$ by $A_2$. At the end the certificate is used in a postprocessing step of the algorithm to decide $k$-connectivity of $G$ without any further input.

Since the certificates are to be memorized by the algorithms they have to be sparse, i.e., consist only of a linear number of edges. There is a lower bound of $kn/2$ on the number of edges in a certificate for a $k$-connected graph. That follows from the fact that in such a graph each vertex has to be of degree at least $k$. So a certificate $G'$ for a $k$-vertex connected graph has minimum degree $\delta(G')\ge k$ and  therefore consists of at least $kn/2$ edges.\\
\indent It is easy to construct a certificate with a minimum number of edges for $1$-connectivity which is just a spanning tree. This can be done in the semi-streaming model in one pass and per-edge processing time of ${\cal O}(\alpha(n))$ \cite{FeigenbaumKannanMcGregorSuriZhang04}. But in general we cannot go for certificates with a minimum number of edges, this problem is ${\cal NP}$-complete \cite{GareyJohnson}. Even for $k=2$ computing a minimum certificate for $k$-connectivity of a graph $G$ is ${\cal NP}$-complete since it tells us if $G$ contains a hamiltonian circuit.\\
\indent To our purposes it suffices to generate certificates not of minimum but of linear size. There are several approaches for this aim. In \cite{NagamochiIbaraki} the authors present a linear time algorithm that generates a certificate for $k$-connectivity of linear size. We do not know if this algorithm can be adopted to the semi-streaming model. In \cite{CheriyanThurimella91} it is shown that the sequential execution of $k$ breadth-first searches and a union of their trees yields a sparse certificate for $k$-connectivity. This approach is not suitable for the semi-streaming model since by a result of \cite{FeigenbaumKannanMcGregorSuriZhang05} computing a breadth-first search tree in the semi-streaming model cannot be done in a constant number of passes over the input.\\
\indent In this paper we use different approaches to construct a sparse certificate for $k$-connectivity. Our first algorithm $A_1$ is an adoption of an online algorithm presented in \cite{CheriyanThurimella91} and \cite{CheriyanKaoThurimella93}. The certificate $G'$ is built up by adding an input edge $uv$ to $G'$ if $u$ and $v$ are not $k$-connected in $G'$. The resulting certificate consists of at most $2kn$ edges \cite{CheriyanKaoThurimella93}.\\
\indent Our second algorithm $A_2$ uses the fact that the union of the forests of $k$ sequentially executed scan-first searches yields a certificate for $k$-connectivity \cite{CheriyanKaoThurimella93}. The number of edges in this certificate as a union of forests is at most $k(n-1)$. We utilize this approach for the semi-streaming model by presenting the semi-streaming algorithm $A_2$ that computes a union of $k$ scan-first search forests.

The two algorithms $A_1$ and $A_2$ presented in the next two subsections differ in various ways: They have different per-edge processing time $T$, run a different number of passes $P$ over the input stream and produce different certificates. While for the first algorithm $T(A_1) = {\cal O}(k^2n)$ and it uses only one pass over the input, i.e., $P(A_1)=1$, for the second algorithm we have $T(A_2)={\cal O}(k+\alpha(n))$ and $P(A_2)=k+1$. Furthermore the certificate $C(A_2)$ produced by the second algorithm is more powerful than $C(A_1)$. It can not only be used to test the input graph $G$ for $k$-connectivity as $C(A_1)$ can, but allows to generate all $l$-separators of $G$ where $l<k$. 


\subsection{Slow Edge-Processing, One Pass}
In this subsection we present a semi-streaming algorithm that in one pass goes over the graph stream of an input graph $G$. It uses ${\cal O}(k^2n)$ time to process each input edge and creates a certificate $C(A_1)$ for $k$-vertex connectivity of $G$.

To derive this algorithm $A_1$ we adopt the online algorithm of Cheriyan and Thurimella \cite{CheriyanThurimella91} and Cheriyan, Kao and Thurimella \cite{CheriyanKaoThurimella93} respectively. Their algorithm runs over a graph stream of the input graph $G$ in one pass and constructs a certificate for $k$-connectivity of $G$. We follow their approach but since they consider no memory restrictions we have to make sure that using this approach does not exceed the memory limitations of the semi-streaming model.

The algorithm itself is simple. At the beginning the certificate $C(A_1)$ is empty. For each input edge $uv$ in the input stream we test whether the current certificate $C(A_1)$ as a subgraph of $G$ contains at most $k-1$ vertex-disjoint paths between $u$ and $v$. If so, we add the edge $uv$ to $C(A_1)$, otherwise the certificate remains unchanged and $A_1$ forgets about this edge and examines the next one.

Since our algorithm processes the edges in the way the algorithm in \cite{CheriyanKaoThurimella93} does, we can assert that $A_1$ indeed constructs a certificate for $k$-connectivity of the input graph.

\begin{lemma}[\cite{CheriyanKaoThurimella93}]\label{GKConn->certKConn}
If the input graph $G=(V,E)$ is $k$-connected, then the final certificate $C(A_1)$ is $k$-connected. \qed
\end{lemma}

We claim that we can keep $C(A_1)$ in the memory of $A_1$ to allow testing of the $k$-connectivity of $G$ using $C(A_1)$ in a postprocessing step. For that reason we have to ensure that $C(A_1)$ does not exceed the memory limitations of the semi-streaming model of ${\cal O}(n\cdot\mbox{polylog}\,n)$ bits. We follow the track of \cite{CheriyanKaoThurimella93}, where it is shown that, using a result of Mader \cite{Bolobas}, the number of edges in the certificate is linear.

\begin{lemma}[\cite{CheriyanKaoThurimella93}]\label{cert<2kn}
The final certificate $C(A_1)$ has at most $2kn$ edges.\qed
\end{lemma}

We can now formulate the final theorem for $A_1$ completing this subsection.

\begin{theorem}
Given a graph stream of a graph $G$ as an input, the algorithm $A_1$ constructs a certificate $C(A_1)$ for $k$-connectivity of $G$. While doing this, $A_1$ uses time ${\cal O}(k^2n)$ per edge. After reading all edges, $A_1$ is able to decide $k$-connectivity of $G$ in a postprocessing step. The space used in total is ${\cal O}(n\cdot\mbox{polylog}\,n)$ bits.
\end{theorem}
\proof Since $C(A_1)$ is a subgraph of $G$ it is immediate that if $C(A_1)$ is $k$-connected, $G$ must be $k$-connected as well. Together with Lemma \ref{GKConn->certKConn} it follows that the final $C(A_1)$ is a certificate for $k$-connectivity of $G$.

For each input edge $uv$ $A_1$ has to check whether there are at most $k-1$ vertex-disjoint paths between $u$ and $v$ in $C(A_1)$. This can be done by constructing a flow from $u$ to $v$ with all node capacities set to one. To compute the flow we use the algorithm of Even and Tarjan \cite{EvenTarjan}. It runs within the memory limitations of ${\cal O}(n\cdot\mbox{polylog}\,n)$ bits since the space used by this algorithm is proportional to the number of edges in the current certificate which has linear size by Lemma \ref{cert<2kn}. 

The flow between $u$ and $v$ has to be computed only up to a value of $k$ and the algorithm of Even and Tarjan can do so in time ${\cal O}(\min\{k,n^{1/2}\}m)$ on a graph with $n$ vertices and $m$ edges. With $m={\cal O}(kn)$ in our certificate and $k$ being a constant a time of ${\cal O}(k^2n)$ suffices to test for $k$ vertex-disjoint paths between $u$ and $v$ for every input edge $uv$. 

The final $C(A_1)$ fits in ${\cal O}(n\cdot\mbox{polylog}\,n)$ bits by Lemma \ref{cert<2kn}. Since it is a certificate for $k$-connectivity of $G$, $A_1$ can test the final $C(A_1)$ for $k$-connectivity to decide $k$-connectivity of $G$. This test can be done in a postprocessing step without any further input. To this end we use the $k$-connectivity algorithm of Gabow \cite{Gabow} that runs in time ${\cal O}(n+\min\{k^{5/2}, kn^{3/4}\}kn)$ and, which is more important, uses a space linear in the size of $C(A_1)$. Consequently the postprocessing step does not exceed the memory limitations of ${\cal O}(n\cdot\mbox{polylog}\,n)$ bits either and $A_1$ is indeed a semi-streaming algorithm. 
\qed


\subsection{Fast Edge-Processing, \boldmath{$k+1$} Passes}
In this subsection the semi-streaming algorithm $A_2$ is presented, which runs over the graph stream input for $k+1$ times and examines each edge in time ${\cal O}(k+\alpha(n))$. It constructs a certificate $C(A_2)$ which can finally be used in a postprocessing step to test the input graph $G$ for $k$-connectivity. 

Our algorithm $A_2$ uses the idea of scan-fist search due to Cheriyan, Kao and Thurimella \cite{CheriyanKaoThurimella93}. \emph{Scan-first search} is a way of systematically marking the vertices of a given graph and works as follows. For a connected graph $G$ we begin with one arbitrary starting vertex marked and all other vertices unmarked. On a marked vertex $v$ we can do the main step called a scan of $v$: That is marking all non-marked neighbors of $v$. After that step $v$ is called scanned and there will be no scanning of $v$ again. In that fashion scan-first search iteratively marks all unmarked vertices and scans all unscanned but marked vertices of $G$ until all vertices are scanned. 

Scan-first search in a connected graph $G$ yields a spanning tree $T$ in the following way. At the beginning $T$ is empty. If we scan a vertex $v$ we add to $T$ all edges from $v$ to its unmarked neighbors. Such a tree is called \emph{scan-first search tree}.\\
\indent We can get a consecutive numbering of the vertices by taking the order in which the vertices are scanned. We call such a numbering a \emph{scan-first search numbering}.

For applying scan-first search to unconnected graphs $G$ we can successively perform scan-first search to every connected component of $G$. That produces a scan-first search tree for every connected component, the union of them we call a \emph{scan-first search forest}.

Note that scan-first search is a generalization of both depth-first search and breadth-first search. If there is more than one vertex marked and unscanned and therefore more than one vertex can be chosen to be scanned next, scan-first search can take any of these vertices. By choosing in every step that vertex which was marked most recently scan-first search performs a depth-first search. If that vertex is chosen that has been marked for the longest time scan-first search proceeds in a breadth-first search manner.

The next theorem shows how we will make use of scan-first search forests to obtain a certificate for $k$-connectivity.

\begin{theorem}[\cite{CheriyanKaoThurimella93}]\label{scanFirstSearch=>kConnectivity}
Given an undirected graph $G(V,E)$ with $n$ vertices and a positive integer $k$. For $i=1,2,\ldots,k$ let $F_i$ be the edge set of the scan-first search forest in the graph $G_{i-1}=(V,E\setminus(F_1\cup\ldots\cup F_{i-1}))$. Then $F_1\cup\cdots\cup F_k$ is a certificate for the $k$-connectivity of $G$ and this certificate has at most $k(n-1)$ edges.\qed
\end{theorem}

If we want to apply the scan-first search approach in our algorithm $A_2$, we have to show how scan-first search can be performed in the semi-streaming model.

\begin{lemma}\label{scanFirstSearchInSemiStreaming}
There is a semi-streaming algorithm $X$ that generates a scan-first search forest $F$ of a graph $G$. To this aim $X$ runs over the graph stream of $G$ twice. $X$ processes each edge in time ${\cal O}(\alpha(n))$ in the first pass and in time ${\cal O}(1)$ in the second pass over the input.
\end{lemma}
\proof In the first pass over the input $X$ computes a spanning forest $Z$ of $G$. This can be done using a disjoint set data structure $D$. We start with $n$ singletons, i.e., $n$ sets containing a single vertex each, and $Z=\emptyset$. For every input edge $uv$ $X$ checks if $u$ and $v$ are in different sets of $D$. In that case $uv$ is added to $Z$ and the sets of $u$ and $v$ are joined.\\
\indent $D$ can be maintained in the memory limitations of a semi-streaming algorithm since for every vertex $v$ we only have to memorize a vertex representing the subset containing $v$. Augmented with path compression and union by rank \cite{Tarjan} the operations on $D$ take time ${\cal O}(\alpha(n))$ for each input edge.\\
\indent At the end of the first pass over the input $X$ has constructed $Z$, which, as a spanning forest of $G$, has at most $n-1$ edges and can be stored in ${\cal O}(n\cdot\mbox{polylog}\,n)$ bits.

After reading the input the first time but before reading it again $X$ performs a depth-first search on $Z$ as an intermediate step. The preorder numbering $0,\ldots,n-1$ of the vertices according to this depth-first search on $Z$ yields an order $o:\{0,\ldots,n-1\}\rightarrow V$ of the vertices, let $o(t)$ be the vertex at position $t$ in that order. While building the order $o$ at the same time $X$ can construct $o^{-1}$, that is, for every vertex $v$ $o^{-1}(v)$ being the position of $v$ in the order $o$.\\ 
\indent The depth-first search can be done in time ${\cal O}(n)$, leaving the amortized time for processing an edge in the first run over the graph stream unchanged. To do the depth-first search and to store the order of the vertices and its reverse, surely ${\cal O}(n\cdot\mbox{polylog}\,n)$ bits suffice.

Note that this very order of the vertices can also be produced as a numbering of a certain scan-first search run $R$ of $G$. $R$ starts at vertex $s=o(0)$. In a connected graph $G$ for the order $o$ holds, that for every $0<a\le n-1$ the vertex $u=o(a)$ is adjacent in $G$ to a vertex $v$ such that $v=o(b)$, $b<a$. So $R$ can scan the vertices in order $o$: If $R$ at step $0<a\le n-1$ should scan vertex $u=o(a)$, there must be a vertex $v$ which has been scanned before and is adjacent to $u$. Therefore $u$ is marked and can be scanned in step $a$. For an unconnected graph $G$ this argumentation can be extended to a sequential execution of the connected components.

The aim of $X$ in the second pass over the graph stream is to simulate the scan-first search run $R$ to produce the scan-first search forest $F$ that $R$ would produce on $G$. Before starting the second pass over the input $X$ knows about the sequence $o$ in which $R$ would scan the vertices of $G$ and will, needless to say, make use of this order.\\
\indent At the beginning $F$ is empty. For each input edge $uv$ in the second pass $X$ looks at the positions $o^{-1}(u)$, $o^{-1}(v)$ of the vertices in the order $o$. Without loss of generality, let $o^{-1}(u) < o^{-1}(v)$. If there is no neighbor $w$ of $v$ in $F$ such that $o^{-1}(w) < o^{-1}(v)$ we add the edge $uv$ to $F$. If otherwise there is a neighbor $x$ of $v$ in $F$ with $o^{-1}(x) < o^{-1}(v)$, $X$ compares $o^{-1}(x)$ and $o^{-1}(u)$. In the case that $o^{-1}(u) > o^{-1}(x)$, $X$ leaves $F$ unchanged, forgets about $uv$ and proceeds to the next input edge. If in the contrary $o^{-1}(u) < o^{-1}(x)$ the edge $vx$ is deleted from $F$ and $uv$ is inserted instead. 

After processing the whole graph stream, for each $uv \in F$ the following holds: If $o^{-1}(u) < o^{-1}(v)$, $u$ is the only neighbor of $v$ among all other neighbors of $v$ that are preceding $v$ in the order $o$. Moreover $u$ is preceding all other neighbors of $v$.\\
\indent For this reason $F$ is exactly the scan-first search forest that $R$ would produce on $G$ scanning the vertices in sequence $o$. An edge $uv$, $u$ preceding $v$, is only put in the scan-first search forest of $R$ if, during the scan of $u$, the unmarked neighbor $v$ of $u$ is marked. For all other neighbors $x$ of $v$ succeeding $u$ but preceding $v$ the edge $xv$ is not added to the scan-first search forest since at the time $x$ is scanned, $v$ is already marked. So $o^{-1}(u)<o^{-1}(x)$ for every other neighbor $x$ of $v$.

For the second pass over the graph stream $X$ can perform all necessary operations in time ${\cal O}(1)$ per edge. Since $F$ as a forest consists of at most $n-1$ edges, $X$ can maintain $F$ in ${\cal O}(n\cdot\mbox{polylog}\,n)$ bits. \qed 

Now we can state our main theorem in this subsection on how $A_2$ computes its $C(A_2)$ as a union of the edges of $k$ scan-first search trees.

\begin{theorem}
Given an undirected graph $G(V,E)$ with $n$ vertices and a positive integer $k$. For $i=1,2,\ldots,k$ let $F_i$ be the edge set of the scan-first search forest in the graph $G_{i-1}=(V,E\setminus(F_1\cup\ldots\cup F_{i-1}))$. A semi-streaming algorithm $A_2$ can compute $C(A_2) :=  F_1\cup\cdots\cup F_k$ using $k+1$ passes over the graph stream of $G$ as input, $F_i$ is computed in pass $i+1$. $A_2$ processes each input edge in time ${\cal O}(k+\alpha(n))$. The final $C(A_2)$ can be used to decide $k$-connectivity of $G$ in a postprocessing step.
\end{theorem}
\proof $A_2$ uses $k$ nested instances of the semi-streaming algorithm $X$ of Lemma \ref{scanFirstSearchInSemiStreaming}. Let $X_i$, $1\le i\le k$, be the ith instance of $X$ called by $A_2$. $X_i$ is called at the beginning of pass $i$ and lasts for two passes of $A_2$, i.e., is finished after pass $i+1$ of $A_2$. That way in pass $i$ of $A_2$ $X_i$ is running its first pass while $X_{i-1}$ performs its second pass.

We will show that every $X_i$ computes $F_i$, a scan-first search forest on $G_{i-1}$. It suffices to show how each $X_i$ does not work on the entire graph $G$ but on $G_{i-1}$ with the reduced edge set $E\setminus(F_1\cup\ldots\cup F_{i-1}))$, since we know due to Lemma \ref{scanFirstSearchInSemiStreaming} that each $X_i$ constructs a valid scan-first search forest for the graph that is presented to it.

In the first pass of each $X_i$, $i>1$, $X_i$ does not see the input of $A_2$, that is, does not see the graph stream input edges. It gets edges handed over by $X_{i-1}$ as input edges. (This handing over of edges from $X_{i-1}$ is described in the next paragraph since it corresponds to the handing over from $X_i$ to $X_{i+1}$.) On this input handed over by $X_{i-1}$, $X_i$ computes an odering $o_{i}$ of the vertices that a scan-first search run on these edges in $G$ might generate as a scan-first search order. The first instance $X_1$ gets the original graph stream edges as input edges.

In the second pass of $X_i$ (which is pass $i+1$ of $A_2$) $X_i$ directly processes the input edges of $A_2$. For each input edge $uv$ in the input stream $X_i$ checks whether $uv$ is an edge in $F_j$, the forest that was computed by $X_j$ for all $j<i$. This can be done in time ${\cal O}(k)$ as follows.\\
\indent For each $X_j$ $A_2$ stores the computed forest $F_j$, the order of the vertices $o^{-1}_{j}$ used by $X_j$ to build $F_j$ and for each vertex $v$ the position $o^{-1}_{j}(w)$ of the at most one neighbor $w$ of $v$ in $F_j$ that is preceding $v$ in $o_{j}$. (See Lemma \ref{scanFirstSearchInSemiStreaming} why this is at most one neighbor.) To test if the input edge $uv$ is in $F_j$ $X_i$ looks at $o^{-1}_{j}(u)$ and $o^{-1}_{j}(v)$. Let w.l.o.g. $o^{-1}_{j}(u) < o^{-1}_{j}(v)$. Let $p$ be the position of the at most one neighbor of $v$ in $F_j$ that is preceeding $v$ in $o_{j}$. If $p=o_{i}^{-1}(u)$ the edge is in $F_j$ otherwise $uv$ is not in $F_j$, since $v$ has only one neighbor in $F_j$ preceding $v$. This way $X_i$ can test for every input edge $uv$ if it is part of one $F_j$, $j<i$, in constant time. Since $j<i\le k$ all $F_j$ can be checked for $uv$ in time ${\cal O}(k)$.\\
\indent If $X_i$ finds the input edge $uv$ existing in one of the $F_j$, $j<i$, it skips the edge and proceeds to the next input edge. In this case  no edge is handed over to $X_{i+1}$ that executes its first pass. If otherwise $uv\not\in F_j$ $\forall j<i$, $X_i$ uses $uv$ to build its scan-first search forest $F_i$ as specified in Lemma \ref{scanFirstSearchInSemiStreaming}. If $uv$ is not inserted in $F_i$, it is handed over to $X_{i+1}$. If $uv$ is added to $F_i$ and $xy$ removed from $F_i$ in exchange, $xy$ is handed over to $X_{i+1}$. In the remaining case that $uv$ is added to $F_i$ and no edge is removed from $F_i$ no edge is handed over to $X_{i+1}$. 

Using the described operations $X_i$ computes the scan-first search forest of $G_{i-1}=(V,E\setminus(F_1\cup\ldots\cup F_{i-1}))$. In the first pass of each $X_i$ it does not see the edges in $F_1\cup\cdots\cup F_{i-2}$ since they are skipped by $X_{i-1}$ and not handed over to $X_i$. Moreover $X_{i-1}$ only hands over to $X_i$ those edges that are not in $F_{i-1}$. So in the first pass of each $X_i$ it builds an order $o_i$ of the vertices that a scan-first search run might generate on $G_i$. In the second pass every $X_i$ only uses the input edges $uv \not\in F_1\cup\cdots\cup F_{i-1}$ to build its $F_i$ according to $o_i$. That yields a scan-first search forest due to Lemma \ref{scanFirstSearchInSemiStreaming}.

It remains to show the claimed time bounds and the memory limitations of $A_2$. In the first pass of $A_2$ only $X_1$ is running using a processing time of ${\cal O}(\alpha(n))$ per edge. In pass $i$, $1<i<k+1$, $X_i$ is running in its first pass getting at most one edge handed over from $X_{i-1}$ per input edge of $A_2$ and processing each of these edges in ${\cal O}(\alpha(n))$. In the same pass $i$ $X_{i-1}$ is executing its second pass. While doing so $X_{i-1}$ tests each input edge of $A_2$ for existence in $F_1\cup\cdots\cup F_{i-2}$ in time ${\cal O}(k)$ and building its $F_{i-1}$ in ${\cal O}(1)$ per edge. In the last pass of $A_2$ only $X_k$ is running checking the input edges for existence in the previous forests and building its scan-first search forest according to $o_k$  in time ${\cal O}(k)$ per input edge.\\
\indent Since each $X_i$ uses ${\cal O}(n\cdot\mbox{polylog}\,n)$ bits a constant number of $k$ instances can not overrun the memory boundaries of the semi-streaming model. Storing the - at most one - neighbor of a vertex preceding this vertex for each $F_i$ does not violate these boundaries either.

After $k+1$ passes each $X_i$ computed its $F_i$ and $A_2$ can merge these $F_i$ producing $C(A_2)$ as a certificate for $k$-connectivity of $G$. In a postprocessing step $A_2$ can use $C(A_2)$ to test $G$ for $k$-connectivity executing on $C(A_2)$ the $k$-connectivity algorithm of Gabow \cite{Gabow} that runs in time ${\cal O}(n+\min\{k^{5/2}, kn^{3/4}\}kn)$ and using linear space in the size of $C(A_2)$.
\qed

The certificate $C(A_2)$ produced by $A_2$ is more powerful than the one of $A_1$. Cheriyan, Kao and Thurimella \cite{CheriyanKaoThurimella93} strengthened  Theorem \ref{scanFirstSearch=>kConnectivity} in the following way. 

\begin{theorem}[\cite{CheriyanKaoThurimella93}]\label{sameSeparators}
Given an undirected graph $G(V,E)$ with $n$ vertices and a positive integer $k$. For $i=1,2,\ldots,k$ let $F_i$ be the edge set of the scan-first search forest in the graph $G_{i-1}=(V,E\setminus(F_1\cup\ldots\cup F_{i-1}))$. Then $G_k=(V,F_1\cup\cdots\cup F_k)$ and $G$ have the same $l$-separators for all $l<k$. \qed
\end{theorem}

So $A_2$ can use its computed certificate $C(A_2)$ in a postprocessing step to identify all $l$-separators for any $l<k$ of the given graph $G$: For every pair of vertices $u,v$ $A_2$ runs the algorithm of Even and Tarjan \cite{EvenTarjan} to determine if there are at most $l<k$ vertex-disjoint paths between $u$ and $v$. If so, any set consisting of one internal vertex of every of these $l$ paths is an $l$-separator in $C(A_2)$ and thus in $G$ by Theorem \ref{sameSeparators}. The space used by the algorithm of \cite{EvenTarjan} is linear in the number of edges in $C(A_2)$ and every $l$-separator with $l<k$ is of constant size. Therefore all $l$-separators, $l<k$, of a given graph can be computed and memorized in a postprocessing step of $A_2$ without exceeding the boundaries of the semi-streaming model.


\section{Conclusions and Open Questions}

We extended the possibility of testing graph $k$-connectivity in the semi-streaming model from $k\le 4$ to $k$ being an arbitrary constant. To this aim we presented two semi-streaming algorithms, both of them computing a sparse certificate for $k$-connectivity of the input graph $G$. In a postprocessing step each algorithm can use its constructed certificate to decide $k$-connectivity of $G$ without exceeding the limits of the semi-streaming model. The second algorithm can use its certificate to generate all $l$-separators for all $l<k$.\\
\indent Due to the memory limitations of the semi-streaming model our approaches cannot be applied for non-constant $k$. We do not know if $k$-connectivity for non-constant $k$ can be determined in the semi-streaming model and what approaches might be suitable.\\
\indent For some real-world applications it is not feasible to allow more than one pass over the input stream. Therefore it is desirable to combine the fast per-edge processing time of our second algorithm with the modesty of our first algorithm reading the input only once.


\section{Acknowledgments}

We are grateful to Stefan Hougardy for helpful discussions.



\begin{thebibliography}{99}

\bibitem{AlonMatiasSzegedy}
N. Alon, Y. Matias, and M. Szegedy. 
The space complexity of approximating the frequency moments. 
In: Proc. 28th ACM STOC, pages 20--29, May 1996.

\bibitem{Bolobas}
B. Bollob\'as.
Extremal Graph Theory. 
Academic Press, London, 1978.

\bibitem{CheriyanThurimella91} 
J. Cheriyan and R. Thurimella.
Algorithms for Parallel $k$-Vertex Connectivity and Sparse Certificates.
In: Proceedings of the 23rd Annual ACM Symposium on Theory of Computing, 1991, pp. 391-401.

\bibitem{CheriyanKaoThurimella93}
J. Cheriyan, M.-Y. Kao, and R. Thurimella.
Scan-first search and sparse certificates: An improved parallel algorithm for k-vertex connectivity.
SIAM J. Comput., 22 (1993), pp. 157--174.

\bibitem{EppsteinGalilItalianoNissenzweig}
D. Eppstein, Z. Galil, G. F. Italiano, and A. Nissenzweig. 
Sparsification - A technique for speeding up dynamic graph algorithms. 
Journal of ACM, 44(1):669--696, 1997.

\bibitem{EvenTarjan}
S. Even and R. E. Tarjan. 
Network flow and testing graph connectivity. 
SIAM J. Computing 4 (1975), 507--518.

\bibitem{FeigenbaumKannanMcGregorSuriZhang04}
J. Feigenbaum, S. Kannan, A. McGregor, S. Suri, and J. Zhang. 
On graph problems in a semi-streaming model.
ICALP 2004, In: LNCS 3142, 531-543, 2004.

\bibitem{FeigenbaumKannanMcGregorSuriZhang05}
J. Feigenbaum, S. Kannan, A. McGregor, S. Suri, and J. Zhang. 
Graph Distances in the Streaming Model: the Value of Space. 
SODA 2005: 745-754.

\bibitem{Gabow}
H. N. Gabow.
Using expander graphs to find vertex connectivity. 
In: Proceedings of the 41st IEEE Symposium on Foundations of Computer Science, IEEE Computer Society, Los Alamitos, CA, 2000, pp. 410--420.

\bibitem{GareyJohnson} 
M.R. Garey, D.S. Johnson.
Computers and Intractability - A Guide to the Theory of NP-Completeness.
W.H.Freeman and Company, 1979.

\bibitem{GilbertKotidisMuthukrishnan}
A. C. Gilbert, Y. Kotidis, S. Muthukrishnan, and M. J. Strauss. 
Surfing wavelets on streams: one-pass summaries for approximate aggregate queries. 
In: Proceedings of the 27th International conference on Very Large Data Bases, 79-88, 2001.

\bibitem{GuhaKoudasShim}
S. Guha and N. Koudas. 
Data-Streams and Histograms. 
In: Proc. 33th ACM Symposium on Theory of Computing, 471-475, 2001.

\bibitem{HenzingerRaghavanRajagopalan}
M. R. Henzinger, P. Raghavan, and S. Rajagopalan.
Computing on data streams.
In: External Memory Algorithms,
Dimacs Series In Discrete Mathematics And Theoretical Computer Science, 50:107-118, 1999

\bibitem{McGregor}
A. McGregor.
Finding Graph Matchings in Data Streams.
APPROX and RANDOM 2005, In: LNCS 3624, 170-181, 2005.

\bibitem{Muthukrishnan}
S. Muthukrishnan.
Data streams: Algorithms and applications.
2003. Available at http://athos.rutgers.edu/$\sim$muthu/stream-1-1.ps

\bibitem{NagamochiIbaraki}
N. Nagamochi and T. Ibaraki.
Linear time algorithms for finding a sparse k-connected spanning subgraph of a k-connected graph. 
Algorithmica, 7:583--596, 1992.

\bibitem{Schrijver}
A. Schrijver. 
Combinatorial optimization: polyhedra and efficiency. 
Springer, Heidelberg, 2003.

\bibitem{Tarjan}
R.E. Tarjan. 
Data Structures and Network Algorithms. 
CBMS-NSF Regional Conference Series in Applied Mathematics, 1983.

\end{thebibliography}
\end{document}